**Full Title:**

Sensory fusion in *Physarum polycephalum* and implementing multi-sensory functional computation

**Authors:**

James G.H. Whiting [1], Ben P.J. de Lacy Costello [1,2], Andrew Adamatzky [1].

[1] Unconventional Computing Centre, University of the West of England, Bristol, UK.

[2] Institute of Biosensing Technology, University of the West of England, Bristol, UK.

**Corresponding author:**

James G. H. Whiting.

Phone number: 01173282461

Email Address: James.whiting@uwe.ac.uk

Department of Computer Science and Creative Technologies

Frenchay Campus

University of the West of England,

Bristol,

UK.

**Abstract:**

Surface electrical potential and observational growth recordings were made of a protoplasmic tube of the slime mould *Physarum polycephalum* in response to a multitude of stimuli with regards to sensory fusion or multisensory integration. Each stimulus was tested alone and in combination in order to evaluate for the first time the effect that multiple stimuli have on the frequency of streaming oscillation. White light caused a decrease in frequency whilst increasing the temperature and applying a food source in the form of oat flakes both increased the frequency. Simultaneously stimulating *P. polycephalum* with light and oat flake produced no net change in frequency, while combined light and heat stimuli showed an increase in frequency smaller than that observed for heat alone. When the two positive stimuli, oat flakes and heat, were combined, there was a net increase in frequency similar to the cumulative increases caused by the individual stimuli. Boolean logic gates were derived from the measured frequency change.

## 1. Introduction:

*1.1. Organism introduction*

The myxomycete *Physarum polycephalum* is a single celled true slime mould which can grow to several centimetres in diameter, living on decaying wood and digesting organic matter such as bacteria and decomposing vegetation. The yellow cellular plasmodium consists of a multinuclear cytoplasm encompassed by a large single membrane, it's shape is determined by the local environment; combinations of nearby chemical sources, light, temperature and other factors affect the direction and shape which *P. polycephalum* grows (Carlilie, 1970; Durham and Ridgway, 1976; Knowles and Carlilie, 1978; Nakagaki et al., 2000; Nakagaki et al., 2004). The organism has been shown to find food sources in mazes as well as being used for binary logic computation, shortest path estimation, solving geometric problems and approximating man-made transport networks (Adamatzky, 2010a; Adamatzky, 2012; Latty and Beekman, 2009; Nakagaki et al., 2001; Nakagaki and Guy, 2008; Shirakawa and Gunji, 2010).

When attractants and repellents are in close proximity to *P. polycephalum*, the cell builds a spatial network of protoplasmic tubes. The cell's protoplasmic tube network connects to sources of attractants along the shortest path while avoiding repellent sources. Some attractants are indicators of potential food sources (Knowles and Carlile, 1978) however there are some chemicals that are chemoattractants but not food sources, instead they could be signalling chemicals from plants and insects in *P. polycephalum's* natural environment (Costello and Adamatzky, 2013; Whiting et al., 2014). Alternatively these chemicals could act directly on the motor function or other key metabolic pathways within *P. polycephalum*. There are also chemicals which act as repellents, which the organism grows away from (Carlilie, 1970; Knowles and Carlile, 1978).

Exposing *P. polycephalum* to light invokes a negative response, causing the cell to migrate away from the source of light (Adamatzky, 2013a). *Physarum polycephalum* shows strong photoavoidance to white (Nakagaki et al., 1999; Wohlfarth-Bottermann and Block, 1981), ultra-violet (Nakagaki and Ueda, 1996) and blue light(Block and Wohlfarth-Bottermann, 1981). Light is also known to initiate the sporulation phase in the organism (Daniel and Rusch, 1962; Golderer et al., 2001; Kroneder, 1999; Takagi and Ueda, 2010), so *P. polycephalum* is capable of detecting different frequencies and intensity of light. Studies reported that warmer temperatures increased the frequency of shuttle streaming oscillations (Kamiya, 1959; Tauc, 1954). Durham and Ridgway observed migration of *P. polycephalum* towards a warmer portion of an agar substrate, with growth occurring along the temperature gradient from 20 to 34$^{o}$C (Durham and Ridgway, 1976). However, no electrical potential recordings had previously been made when heat was employed as a stimulus. *P. polycephalum* is evidently able to survive in a wide range of temperatures in the wild, indeed the authors have maintained cultures and observed growth from 2$^{o}$C to 35$^{o}$C. *P. polycephalum* is often incubated and experimented with at between 20 to 35$^{o}$C (Coggin and Pazun, 1996; Durham and Ridgway, 1976). Nakagaki demonstrated protoplasmic vein formation due to generation of regular oscillations when applying transient increases of 10$^{o}$C above room temperature (Nakagaki et al., 1996; Nakagaki et al., 2000). Since these papers have noted the increase in oscillation frequency when *P. polycephalum* was warmed, it was decided that electrical recording of a protoplasmic tube should be performed and would constitute one stimulus while investigating sensory fusion of *P. polycephalum*.

It is not understood how *P. polycephalum* combines data from several opposing or additive stimuli. It is well documented that *P. polycephalum* responds to different environmental stimuli as described above, however there are several attractant and repellent stimuli present simultaneously within the natural environment, and how *P. polycephalum* computes, prioritises and acts when exposed to these various stimuli is totally unknown. The slime mould moves and grows by employing protoplasmic shuttle streaming, where waves of contraction and relaxation of the membrane forces cytoplasm to flow through protoplasmic tubes, extending the tip towards an attractant; it is thought that repellents inhibit this streaming. Time-lapse photography and video microscopy has shown the forward and backward flow of cytoplasm; the oscillation occurs with a period of between 50 to 200 seconds. This oscillation has also been measured electrically (Adamatzky and Jones, 2011; Whiting et al., 2014) and the frequency correlates approximately with that measured optically.

*1.2. Known stimuli in Physarum polycephalum*

Basic carbohydrates are the most tested chemotactic substances, with glucose, galactose and manose being the strongest attractants of those investigated in the literature (Carlilie, 1970; Kincaid and Mansour, 1978b; Knowles and Carlile, 1978). Often a culture of *P. polycephalum* is fed with oat flakes, which are approximately 70% carbohydrate. It has also been demonstrated that Agar itself is a chemo-attractant, with 8% concentrations of agar gel producing stronger responses than oat flakes by way of increased frequency (Whiting et al., 2014). Non-food sources also have chemotactic properties with enzyme cyclic 3',5'-AMP phosphodiesterase inhibitors among the compounds tested (Kincaid and Mansour, 1979), additionally volatile organic chemicals showed chemotactic properties (Costello and Adamatzky, 2013; Whiting et al., 2014). While all the studies of attractant and repellent stimuli on *P. polycehpalum* have been performed individually, there is no study detailing the effect on the growth or frequency of streaming in the organism while changing multiple variables, but there is presumably a basic hierarchy of stimuli, which controls *P. polycephalum's* behaviour. This study documents the effects of combinations of multiple stimuli on the oscillation frequency and spatial growth of *Physarum polycephalum*.

*1.3. Multisensory integration*

It is well known that other organisms combine several senses; every organism which can sense its environment must perform some kind of sensory fusion, to understand its surroundings and plot resultant behaviour. Multisensory integration, or sensory fusion, in animals more developed than sponges is performed by a nervous system (Calvert et al.,

2004), while simple or unicellular organisms have nervous system homologs of intra- and inter- cellular communication pathways (Sakarya et al., 2007). Prokaryotes have a system of biochemical processes which govern behaviour; stimuli driven protein transmitter and receptor driven responses (Bourret, 2006), often having multi-component levels (Bren and Eisenbach, 2000) allowing for competitive or inhibitive binding on receptor sites which, when responding to multiple stimuli, is akin to multisensory integration (Clarke and Voigt, 2011). Eukaryotic fungi have also demonstrated multi component stimuli response signalling pathways through genome analysis (Catlett et al., 2003). It is likely that *Physarum polycephalum* uses multi-component pathways to perform multisensory integration; chemotaxis, phototaxis thermotaxis and other stimuli are combined to change parameters of protoplasmic streaming which ultimately governs the direction of movement and life cycle stages.

*1.4. Logic gates*

Boolean logic is a field of mathematics which describes binary arithmetic, calculating truth values for logic inputs; a logic gate is the physical manifestation of a Boolean operation. Boolean logic is fundamental to computer science as electronic logic gates form the basis of digital operations in computers. The logic operations AND, OR and NOT are the basic operations of Boolean logic; combinations of these basic functions can produced derived gates such as XOR or more complex multi-gate logic. Modern digital logic gates are made up of discrete electronic components such as transistors and resistors, which can be miniaturised and operate at fractions of a second. Organism based logic gates have also been attempted, scientists have implemented a Boolean NOR gate in *Escherichia coli* using fluorescence expression (Tamsir et al., 2011). It has been shown that bacteria inherently calculate Boolean logic during DNA transcription (Bonnet et al., 2013). Yeast too, show AND-gate like control of DNA promoters (Teo and Chang, 2014); even mammalian cells have multi-input logic control of gene regulation which has been adapted into Bio-Logic gates (Kramer et al., 2004). *P. polycephalum* is also capable of computing Boolean logic AND, OR and NOT gates (Tsuda et al., 2004) based on growth along agar channels towards attractants. Other papers have added to this field with cascaded *Physarum polycephalum* gates (Adamatzky, 2010b; Jones and Adamatzky, 2010; Schumann and Adamatzky, 2011) producing more complex combinational logic such as a one-bit adder. This paper aims to investigate sensory fusion in *P. polycephalum* and implement various computational functions.

**2. Method:**

*2.1. Electrical measurement of Physarum polycephalum*

*Physarum polycephalum* was grown in accordance with the common agar culture method, described in detail in a previous paper (Whiting et al., 2014). For experimentation the protoplasmic tube was grown on a customised 9cm petri dish, similar to the setup we showed previously (Adamatzky, 2013a; Whiting et al., 2014) with two 50mm long strips of 10mm wide electrically conductive aluminium tape (Farnell, UK) aligned in a petri-dish with a 10mm gap in the middle; 1ml 2% non-nutrient agar drops were placed on the tape tips as demonstrated in figure 1 to form hemisphere electrodes.

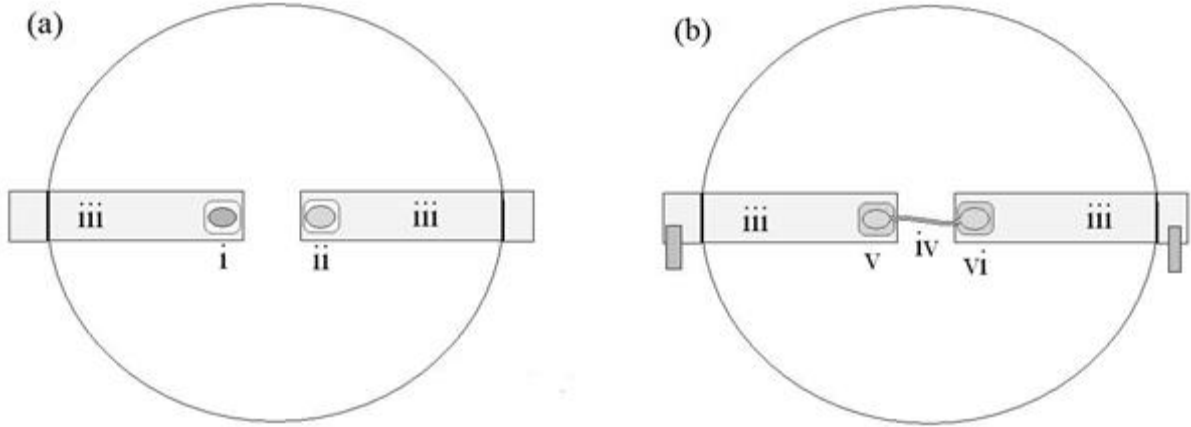

Figure 1. (a) Petri dish set up for stimuli assessment before growth. (b) Petri dish with successful protoplasmic tube formation, connected for electrical potential recording. (i) Bare oat flake. (ii) *Physarum polycephalum* inoculated oat flake. (iii) Conductive tape. (iv) Single protoplasmic tube between electrodes. (v) Positive recording electrode. (vi) Ground reference terminal.

A *P. polycephlum* inoculated oat flake was placed on the ground electrode and a bare oat flake was placed on the recording electrode; over the course of several hours to days a protoplasmic tube would grow across the gap to digest the bare oat flake, when this had occurred the Petri dish was ready for recording. Electrical measurement of *P. polycephalum* was performed on the tube between the two agar electrodes by connecting the ends of the conductive tape to a PicoLog ADC-24 high resolution analogue-to-digital data logger (Pico Technology, UK); this was connected via USB to a laptop installed with PicoLog Recorder software to record the data. The PicoLog ADC-24 recorded ±39 millivolts at 1 hertz for the duration of the experiment, with a 24 bit resolution.

*2.2. Determining frequency change with single or multiple stimuli*

There were three main stimuli types in this experiment, Light, Heat and Chemical, all previously shown to produce taxis of *P. polycephalum*. In order to determine the combined effects of multiple stimuli, it was necessary to measure the range of response to each stimulus individually as detailed in the following paragraphs. Recording was initiated when the tubes had been formed and maintained until completion, the real-time output of the tube was observed and after 10 minutes of oscillatory output was shown with no visible spikes or noise, the start of stimulation (stimulation point) was then begun and maintained for 3 hours, while the recording was only maintained for 10 minutes. The initial 10 minute period before the stimulation was termed the pre-stimulation period, the subsequent 10 minutes were termed the post-stimulation period. The frequency of the pre-stimulation ($f_{pre}$) and post-stimulation ($f_{post}$) phases were calculated using a custom script Fast Fourier Transform in Mathworks Matlab R2012a software. The percentage of frequency change was calculated using equation 1, and expressed as a percentage change. The long term effects of the stimuli on growth were recorded after 3 hours of maintained stimulation. For each set of stimuli 12 successful recordings were made to allow statistical analysis.

$$frequency\ change = {f_{post}}/{f_{pre}} \qquad \text{(Equation 1)}$$

12 control recordings were made where no stimulus was applied. Chemical stimuli in the form of oat flakes has been very well documented and has shown very strong and repeatable positive chemotaxis (Adamatzky, 2010a; Nakagaki et al., 2001; Nakagaki et al., 2007a). The addition of a single oat flake to the colonised agar on top of the recording electrode constituted chemical stimulation. For optical stimulus, the recording electrode was exposed to white light from a 34 watt halogen bulb placed 50 centimetres above the Petri dish while the ground electrode and protoplasmic tube remained in darkness. In preliminary tests of the same set up, a digital thermometer was used to test the change in temperature of the recording electrode agar, there was occasionally a minor increase in temperature (+ 0.5$^o$C max) in the vicinity of the recording electrode caused by exposure to the light. This minimal rise in temperature was deemed not to affect the behaviour of *Physarum polycephalum* as daily temperature fluctuations were of similar magnitude. A 1.4 Watt Peltier element (PE) (RS Components Ltd, UK) 8mm x 8mm x 3.6mm in size was placed under the recording electrode below the Petri dish and the supply voltage and current was calibrated such that it heated up the agar substrate by 10$^o$C above room temperature (room temperature was between 18-21$^o$C). Switching on the Peltier device was performed at the stimulation point; the recording electrode was heated by 10$^o$C in less than 30 seconds.

*2.3. Sensory combinations*

White light and an oat flake were used as dual stimuli to investigate the sensory fusion of these stimuli and subsequently test the power of food sources over light; as known from the literature, oat flakes are strong attractants (Adamatzky, 2010b; Carlilie, 1970; Latty and Beekman, 2009; Nakagaki et al., 2007a; Nakagaki and Guy, 2008; Tsuda et al., 2004) while light is an identified repellent (Nakagaki et al., 1999). White light stimulus was applied to the recording electrode using the method described above while simultaneously adding an oat flake. Increased temperature and white light were tested together using the individual methods simultaneously. Heat and light were applied simultaneously using the methods described in the individual applications, likewise heat and oat flake stimuli.

## 3. Results:

Table 1. Summary of stimuli type and the resultant change in frequency.

| Stimuli Type | Median frequency change | Standard deviation |
|---|---|---|
| Control | 2.1% | 6.9 |
| Oat only | 12.2% | 12.6 |
| Light only | -12.5% | 6.5 |
| Heat only | 19.8% | 8.8 |
| Heat and Light | 14.4% | 12.5 |
| Light and Oat | -1.5% | 11.2 |
| Heat and Oat | 33.2% | 9.6 |

*3.1. Individual stimuli*

Frequency change was calculated to evaluate the effect on protoplasmic streaming oscillations (table 1), while prolonged observations were maintained. For the control group (n=12), the frequency change was a median of 2.1% with standard deviation (SD) 6.9 which indicates there is a very small median change in frequency with no stimulus applied however the inter-quartile range was 10.1% suggesting there is some natural deviation in frequency over time. Chemical stimulus in the form of an oat flake was shown to increase the frequency with a median change of 12.5% (SD 12.5) which is significantly different to the control group with a confidence interval greater than 99% when using the Man-Whitney Test for independent non-parametric variables; this test was used to compare all the following variables. The growth observation for an oat flake showed colonisation in 12 of 12 Petri dishes confirming that an oat flake is a reliable attractant. White light showed a decrease in frequency with a median change of -12.5% (SD 6.5). The decrease is statistically significant with a confidence interval greater than 99%. After 3 hours, *P. polycephalum* completely retreated from the recording electrode in all 12 cases, suggesting white light is a reliable repellent. An increase in recording electrode temperature by 10$^o$C was followed by a median frequency change of 19.8% (SD 8.8) indicating that it is a strong attractant. After an hour of increased temperature, *Physarum polycephalum* had completely migrated to the warmer recording electrode in 5 out of 12 cases while the other 7 remained connected with a protoplasmic tube; extended temperature increase eventually produced sclerotia formation on the recording electrode in all 12 cases due to the dehydration of the agar as a result of the warm conditions.

*3.2. Combined stimuli*

The positive oat flake stimulus appeared to be cancelled out by negative white light exposure; the frequency change was -1.5% (SD 11.2). The median frequency change was slightly lower than the control, however as it had a very similar distribution as shown in figure 2, there was no statistical difference between these groups. For the prolonged observation, 10 of 12 times *P. polycephalum* retreated back to the ground referenced electrode and did not colonise the oat flake, another occasion showed *P. polycephalum* colonising the oat flake, with the remaining instance *P. polycephalum* continuing to occupy the recording electrode yet not colonising the oat flake.

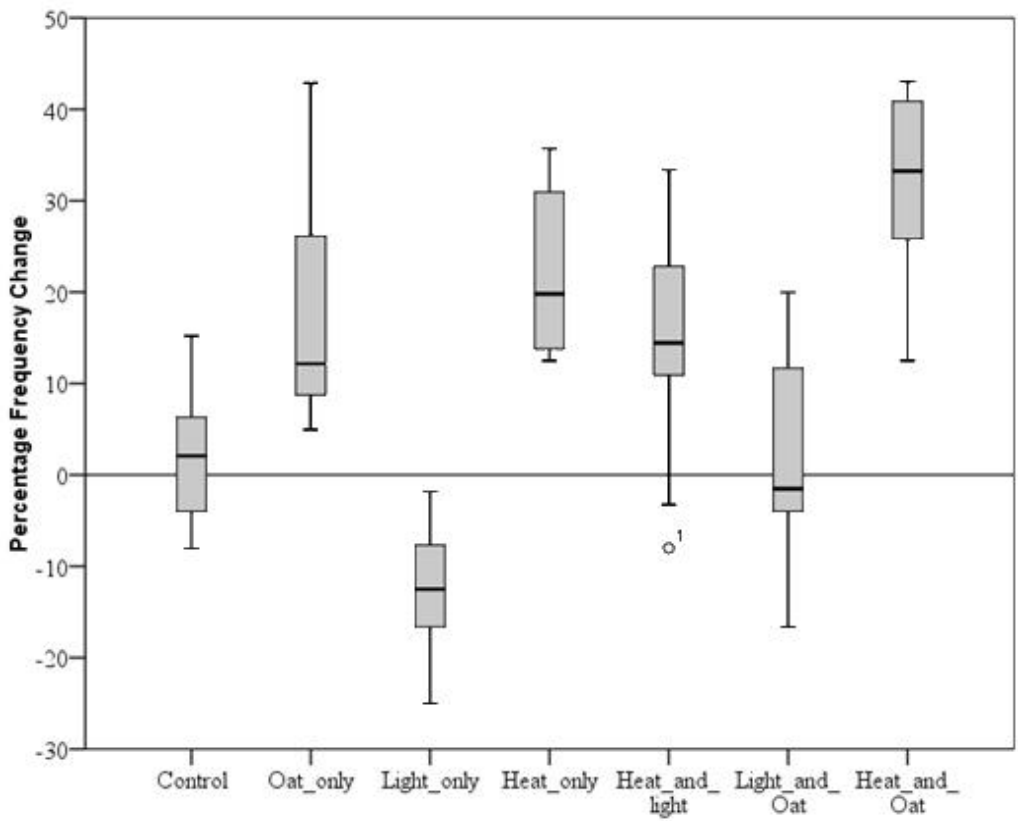

Figure 2. A Box and Whisker plot of frequency changes when testing different stimuli.

When heat and light were both applied the median frequency change was 14.4 (SD 12.5), with 1 outlier marked on figure 2; while there was an increase in frequency in 83% of cases, it was not as marked as the increase when heat alone was applied. The effect from heat and light was significantly different to the control group however it was not significantly different to the effect of heat alone. In 6 of 12 observed cases, *P. polycephalum* remained on the recording electrode, in 2 cases it grew over the Petri dish away from the light and in the remaining 4 cases it retreated back to the ground referenced electrode. The median frequency change when both heat and an oat flake were applied was 33.2 (SD 9.6). The change was statistically significant (to a confidence interval of 95% (*p*-value 0.027)) when compared to the increase measured when only heat was applied., In addition the measured change was statistically significant to all the other groups with a confidence interval greater than 99%. After a matter of hours, *P. polycephalum* had fully inoculated the oat flake in 11 of 12 cases and in the remaining case *P. polycephalum* had grown protoplasmic tubes out from the recording electrode without colonising the oat flake.

**4. Discussion:**

*4.1. Individual stimulus observations and electrical response*

The positive chemical stimulation in the form of an oat flake increases the frequency of every repeat performed with a median frequency increase of 12.2%; when comparing this to the volatile organic chemicals (VOCs) tested previously (Whiting et al., 2014), shows it to be not as strong an attractant as farnesene (frequency increase of 22.5%) and very similar to tridecane (frequency increase of 17%). Farnesene is not a known food source of *P. polycephalum* (Costello and Adamatzky, 2013), demonstrating that although food sources are strong attractants, there may be other chemicals that elicit a stronger attractant response. There may be many reasons why non-food sources have an effect; they may have a direct effect on the motor control pathways, altering protoplasmic streaming. Some previously tested non-food sources are non-selective phosphodiesterase inhibitors, which could increase intracellular cAMP levels, linked with control of *P. polycephalum* movement (Matveeva et al., 2011). It has been shown that there is not always a correlation between nutritional benefit and chemo-attractive strength (Kincaid and Mansour, 1978b). For 12 of 12 occasions *P. polycephalum* colonised the oat flake within a number of hours which confirms the reliability of food sources for maintaining culture.

Exposure to white light showed a marked and statistically significant decrease in frequency by 12.5%, suggesting that white light acts to repel *P. polycephalum* by decreasing the frequency of shuttle streaming oscillation. This confirms previous findings that light sources are repellent, however Adamatzky (Adamatzky, 2013a) reported that white light marginally increases the frequency of oscillation by 8% but notes the large standard deviation and the variable effect of wavelength of light on the frequency and that there was no distinction between green and white light by *P. polycephalum*. Retraction from recording electrode in every case demonstrates the repellent power of light; it has been reported that if *P. polycephalum* is unable to avoid light sources it either fragments (Kakiuchi et al., 2001) or induces sporulation (Daniel and Rusch, 1962; Golderer et al., 2001; Kroneder, 1999; Takagi and Ueda, 2010) however it appears that if *P. polycephalum* is partially illuminated, it will move into the darker area (Adamatzky, 2010a) which is confirmed by these findings. The wavelength of the light appears to affect the frequency change (Adamatzky, 2013a) so it is possible that effects are stronger for different colours with a spectrum of frequencies causing different behaviour; while there is a large spectrum of light which P. polycephalum can detect, it is possible that more complex computing is possible by employing several different or dynamic wavelengths of light. Previous studies reported that response to intensity of light is marginally linear (Schreckenbach et al., 1981; Wohlfarth-Bottermann and Block, 1981), with very little response to low levels of light while increasing the intensity of the light source forces a linear change in frequency; this research agrees mostly, with low levels of light presenting no statistical change during preliminary set-up experiments however further work may be required to correlate intensity of light with electrically measured frequency change.

An increase of $10^{o}$C above ambient temperature increased the frequency of oscillation by approximately 20%. It has been known for a while that *P. polycephalums'* optically measured shuttle steaming frequency changes with the application of heat (Kamiya, 1959; Tauc, 1954), also known is the attractant qualities of heat sources (Durham and Ridgway, 1976), therefore the findings of this test confirms that the electrical oscillation is correlated to shuttle streaming. After several hours *P. polycephalum* migrated towards the source of the heat, however due to prolonged warming and drying out of the agar, sclerotia had formed probably a direct result of plasmodia dehydration. The organism chose to remain in the warm place and form sclerotia despite the ongoing dehydration rather than migrating away from the source of the heat; it may not be biologically disadvantageous to form sclerotia when no other positive stimuli are present in the local environment. When developing the experiment, different levels of heat were applied to the agar, ranging from $5^{o}$C increase to $15^{o}$C increase however there was no statistical change in frequency or behaviour compared to the control when the agar electrode was warmed by $5^{o}$C while an increase of $15^{o}$C caused the agar to dehydrate very quickly resulting in total cessation of the measured oscillations in 80% of cases.

While exposure to white light clearly forces photo-avoidance behaviour, the two positive stimuli (oat flake and warmth) appear to invoke different magnitudes of response; the median frequency changes for oat flakes and heat are 12.2% and 19.8% respectively. There is marginal statistical significance between the two stimuli (*p*-value=0.08 for a 1 tailed test) although there is a trend for heat to increase the frequency to a greater extent than oat flake addition. When developing a *P. polycephalum* computer it would be useful to be able to provide a wide range of reliable and stepped frequency responses, therefore it is possible that a weaker chemotactic chemical, a smaller volume of oat flake or even placing the oat further away from the recording electrode (Nakagaki et al., 2007b) would provide a statistically significant increase in frequency with a magnitude half way between heat and no stimulus; the latter by way of decreasing the chemical gradient which has previously been shown to reduce the response from *P. polycephalum* (Carlilie, 1970; Durham and Ridgway, 1976). Heat could also be altered to provide stepwise or linear response to stimuli in order to develop more complex slime mould gates or analogue controllers.

*4.2. Combination of stimuli*

When an attractant and repellent stimulus are combined there is a short term net effect on the surface electrical potential, cancelling each other out when the two stimuli are equally balanced and opposite. When light and oat are combined in the same experiment there is a median frequency change of -1.5, showing a slight decrease in frequency; there is no statistically significant difference between this and the control group. While the electrical effect appears to be totally cancelled out, the long term migration observation shows migration away from the food source in 83% of cases, suggesting there is an inherent dislike of light by *P. polycephalum*, even with the presence of a an oat flake. It would appear that the repellent effect of light is stronger than the attractant power of a food source in this set-up.

When the negative stimulus light is combined with the positive stimulus heat, there is an increase in frequency which is less than that when heat alone is applied; the combined effect is similar to the effect of heat alone (*p*-value=0.109 with a one tailed test). The percentage increase of frequency when heat is applied is approximately 20% however when light is combined with heat, the increase falls to approximately 14%. In the observed investigations of combined heat and light, *P. polycephalum* remained in place on the recording electrode in half the experiments while avoiding the light in the remaining experiments. The effect of heat on *P. polycephalum* is evidently strongly positive, the magnitude of response is greater than with an oat flake suggesting that heat is the greater of the two stimuli. Light, food and heat are prioritised such that *P.polycephalum* will migrate into illuminated areas so long as they are warm, but not if they contain food sources. It is presumed that heat, at least $10^{o}C$ greater than room temperature, is higher in *P. polycephalum's hierarchy* of movement factors than the presence of an oat flake, however higher temperatures will become inhibitory due to drying and eventual enzyme denaturing. The room temperature at time of experimentation was between 18 and $21^{o}C$, meaning the temperature of *P. polycephalum* during heating was 28 to $31^{o}C$ possibly near the optimal temperature for enzyme or metabolic reactions.

When two positive stimuli are combined, the frequency change is greater than if either stimuli were tested alone; statistically significant compared to either oat only (*p*-value=0.01) and heat only (*p*-value=0.024). The median magnitude of frequency increase when the two stimuli are combined is 33% while the individual median increase due to oat flake and heat are 13% and 20% respectively; this strongly suggests that the effects are purely additive rather than synergistic. It is known that the threshold of chemical detection is affected by temperature (Ueda et al., 1975) with authors implying membrane structure hence detection threshold is temperature dependant so perhaps the pathway by which the combined stimuli is greater is the result of increased chemical detection. This is evidence of complex sensory fusion in the slime mould *P. polycephalum*.

While the stimuli for heat and food are both positive, they are totally different physical stimuli, they may act independently on *P. polycephalum* yet have the same effect of the frequency of shuttle streaming. Light is a negative stimulus and is believed to cause phototaxis with photoreceptors which respond to white or blue light, the latter resembles absorption of flavoprotein (Schreckenbach et al., 1981) which decreases glucose production and decreases frequency of shuttle streaming; the pathway for this is unknown. White light mediates a $H^{+}/K^{+}$ exchange, changing the pH of the plasmodia which induces sporulation however this pathway also remains unknown (Schreckenbach et al., 1981). It has been shown that temperature affects the threshold at which chemoreception occurs (Ueda et al., 1975), while membrane chemoreceptors have been suggested as the most likely method of chemoreception with cyclic nucleotides acting as secondary messengers inside the cell (Aldrich and Daniel, 1982).

$Ca^{2+}$ and cAMP are hypothesised as the biochemical oscillator system which controls shuttle streaming (Smith and Saldana, 1992) and chemoreceptors and photoreceptors may promote or inhibit the cyclic production of $Ca^{2+}$ controlling shuttle streaming. It is possible that oat flakes and other chemotactic chemicals act on specific chemoreceptors in the membrane, however some VOCs would interrupt membrane receptors or act directly on signalling pathways controlling shuttle streaming. As mentioned, temperature affects the chemoreception threshold, perhaps by metabolic or enzymatic effect, therefore the increase in frequency of shuttle streaming when increasing the temperature of the cell, as shown in this paper, is not a behaviour, unlike chemotaxis, but rather a passive change due to temperatures closer to that of the enzymatic optimum.

Potentially all stimuli which exert a form of taxis, whether it negative or positive, are doing so via different pathways acting on the same biochemical oscillator which is responsible for shuttle streaming. This is exemplified by the apparent neutralisation of positive and negative stimuli of light and oat, and to some degree, light and heat.

*4.3. Comparison with chemical data*

When comparing the frequency change of VOCs to that of an oat flake (Whiting et al., 2014), the chemicals linalool and nonanal are stronger repellents than light is; conversely farnesene is a stronger attractant than an oat flake, however the combined effects of an oat flake and heat demonstrates a greater frequency change than farnesene. Volatile organic chemicals diffuse through the air and may act upon *P. polycephalum* membrane receptors; it is unclear how *P. polycephalum* is attracted to oat flakes over a distance, previous studies showed glucose content, as well as other simple carbohydrates, was a strong factor in migration (Carlilie, 1970; Kincaid and Mansour, 1978a; Knowles and Carlile, 1978) though this occurred through agar diffusion rather than through the air. Using specified combinations of stimuli, we can task *P. polycephalum* with computing boolean logic functions, using the frequency change to determine the output. When using oat flake and heat as logic inputs $x$ and $y$, *P. polycephalum* can approximate Boolean logic, as shown in table 2.

Table 2. Theoretical and experimental implementation of logic OR and AND gates, approximated by a combination of two attractant stimuli on *P. polycephalum*.

| Logic input $x$ | Logic input $y$ | OR output | AND output | Median frequency change |
|---|---|---|---|---|
| Oat | Heat | | | |
| 0 | 0 | 0 | 0 | 2% |
| 0 | 1 | 1 | 0 | 20% |
| 1 | 0 | 1 | 0 | 13% |
| 1 | 1 | 1 | 1 | 33% |

*4.4. Implementing logic functions using Physarum polycephalum*

The frequency change when exposed to the heat and oat stimuli is repeatable and of similar magnitude (Fig. 2) we can use this reliable change to approximate Boolean logic with logic 0 and 1 if the frequency change from these stimuli is within certain value ranges. Boolean logic operation OR is implemented with a threshold of 10% increase in frequency while AND uses a threshold of 24% increase in frequency. A NOT gate can be implemented when light is used as an input, with white light representing the input $x$ and using a threshold of -5.5% frequency change. This information is summarised in table 3.

Table 3. Type of logic gate implemented is dependent on thresholds of frequency change, Θ:

    OR gate.    Θ < 10%, Logic 0.    Θ ≥ 10%, Logic 1.

    AND gate.    Θ < 24%, Logic 0.    Θ ≥ 24%, Logic 1.

    NOT gate    Θ < -5.5%, Logic 0.    Θ ≥ -5.5%, logic 1.

The logic OR, AND and NOT gates derived in this paper demonstrate that logic functions in P*hysarum polycephlaum* can be performed accurately, orders of magnitude faster than the growth based logic implementations. Tsuda originally implemented the growth based *P. polycephalum* logic gates with OR, AND and NOT gates giving 100%, 69% and 83% accuracy respectively, values which are comparable to the accuracy of frequency based logic operations presented in this paper (table 4).

Table 4. Accuracy of derived *Physarum polycephalum* logic gates

|  | Logic operation | | |
| --- | --- | --- | --- |
|  | OR | AND | NOT |
| Correct output | 90.3% | 79.2% | 91.7% |

Until now, *P. polycephalum* logic computation used growth and migration (Jones and Adamatzky, 2010; Tsuda et al., 2004; Tsuda et al., 2006). These computations took several hours to complete due to the slow rate of organism growth. Frequency-change based logic implementation is significantly faster than any performed using growth as the calculation, with calculations lasting between 20 and 30 minutes. The main advantage of this electrically recorded implementation is the speed of processing; it has been previously reported that while the migration response of *P. polycephalum* to stimuli is slow with speeds of up to 5cm per hour (Aldrich and Daniel, 1982), electrically recorded responses to chemical, mechanical and optical stimuli are immediate (Adamatzky, 2013a; Adamatzky, 2013b; Adamatzky, 2014; Whiting et al., 2014). The *P. polycephalum* logic gates are designed by interpretation of results where *P. polycephalum* processes inputs and closely describes the output by way of frequency change. OR and AND logic gates may closely be approximated when using the combined effects of Oat and Heat as inputs *x* and *y*, the NOT logic gate uses Light as described in the method section as an input this is tested using obtained data of frequency response using light as a single stimuli. Thresholds are implemented in order to divide the categories of logic 1 and logic 0. With a marginal overlap of the frequency changes of different stimuli, there is some error when approximating the logic outputs; table 3 highlights the accuracy of the logic operations when using the frequency change as a basis for Boolean logic operation and using the threshold, Θ as defined above for each gate. The accuracy was determined by the number of correct outputs using *P. polycephalum's* frequency change, displayed as a percentage, for all 4 input combinations of *x* and *y*; each combination was repeated 12 times for each gate type. A correct output was given when the frequency change as a result of the stimuli fell the correct side of the threshold. This paper has demonstrated short and long term multi-sensory integration in *P. polycephalum*; in addition, advanced frequency change based logic gates were derived using this behaviour, significantly advancing of *P. polycephalum* computing.

## 5. Conclusion:

It has been shown that light exerts negative phototaxis in *Physarum polycephalum*, while an oat flake exerts positive chemotaxis and an increase in temperature of 10$^{o}$C exerts positive thermotaxis. Exposure to white light decreases the measured frequency of a protoplasmic tube by 12.5% while an oat flake increases it by 12.2% and a temperature increase of 10$^{o}$C increases it by 19.8%. This paper shows for the first time the hierarchy of different stimuli, as well as the combined effect of more than one stimuli type; white light on *P. polycephalum* appears to cancel out any effect of oat flake presence while the effect of heat is reduced when white light is shone on the organism. Two positive attractant agents, oat flakes and heat, when combined show an enhanced effect approximately cumulative of the individual effects. The mechanisms of these effects are currently unknown however it is demonstrated that *P. polycephalum* is capable of complex sensory fusion. There is an inherent frequency change of *P. polycephalum* even when no obvious or intended stimulus is present. *Physarum polycephalum* is able to compute logic when placed in certain conditions, its behaviour and frequency changes approximates the Boolean logic functions NOT AND and OR; this has been conceptually proven in this paper. The decrease in computation time over growth based *Physarum polycephalum* gates is significant, completing tasks in less than 30 minutes. Future work could see the development of more advance organism based computation.